# Mispronunciation Detection of Basic Quranic Recitation Rules using Deep Learning

Ahmad Al Harere*, Khloud Al Jallad

Department of Information and Communication Engineering,
Arab International University, Daraa, Syria.

*Corresponding author. E-mail(s): 201710632@aiu.edu.sy; Contributing authors: k-aljallad@aiu.edu.sy;

**Abstract**

In Islam, readers must apply a set of pronunciation rules called Tajweed rules to recite the Quran in the same way that the angel Jibrael taught the Prophet, Muhammad. The traditional process of learning the correct application of these rules requires a human who must have a license and great experience to detect mispronunciation. Due to the increasing number of Muslims around the world, the number of Tajweed teachers is not enough nowadays for daily recitation practice for every Muslim. Therefore, lots of work has been done for automatic Tajweed rules' mispronunciation detection to help readers recite Quran correctly in an easier way and shorter time than traditional learning ways. All previous works have three common problems. First, most of them focused on machine learning algorithms only. Second, they used private datasets with no benchmark to compare with. Third, they did not take into consideration the sequence of input data optimally, although the speech signal is time series. To overcome these problems, we proposed a solution that consists of Mel-Frequency Cepstral Coefficient (MFCC) features with Long Short-Term Memory (LSTM) neural networks which use the time series, to detect mispronunciation in Tajweed rules. In addition, our experiments were performed on a public dataset, the QDAT dataset, which contains more than 1500 voices of the correct and incorrect recitation of three Tajweed rules (Separate stretching[1], Tight Noon[2], and Hide[3]). To the best of our knowledge, the QDAT dataset has not been used by any research paper yet. We compared the performance of the proposed LSTM model with traditional machine learning algorithms used in SoTA. The LSTM model with time series showed clear superiority over traditional machine learning. The accuracy achieved by LSTM on the QDAT dataset was 96%, 95%, and 96% for the three rules (Separate stretching, Tight Noon, and Hide), respectively.

**Keywords:** Deep Learning, Speech Recognition, Mel-Frequency Cepstral Coefficient (MFCC), Ahkam Al-Tajweed, Computer-Assisted Language Learning, Mispronunciation Detection

## 1. Introduction

Due to the importance of speech in human voice communication, Computer-Assisted Language Learning (CALL) systems have gained great importance in the last decade. CALL systems are computer applications that use information technology to teach and learn a new language[1], often not a native language and sometimes referred to as a second language, and to make the learning process much easier, where conventional methods are frequently challenging because of the high cost of learning or there are not enough foreign language teachers available. CALL systems include a Computer-Assisted Pronunciation Training (CAPT) system that diagnoses and identifies mispronunciation and give useful feedback [2], where identifying pronunciation mistakes enables the language learner to obtain an accurate assessment of pronunciation correctness.

---

[1]Separate Stretching: It is the dowel letter at the end of the word and the letter "Hamza" at the beginning of the word that comes after it. There are four or five movements in total [24].
[2] Tight Noon: "Ghunnah" pronunciation requires two movements to be shown in the aggravated "Noon" letter [24].
[3] Hide: The consonant "Noon" letter or "Tanween" is being pronounced for two movements in a non-emphatic state between showing and fading [24].

Mispronunciation detection has received a lot of attention for a variety of languages [3], [4]. However, Arabic has received less attention than other languages [5], although the Arabic language is very important since the Holy Quran was revealed in Arabic and all non-Arab Muslims around the world learn Arabic to complete their religious duties and recite the Quran properly.

The Holy Quran is recited correctly by following a set of pronunciation rules known as the Tajweed rules. Applying these rules means reciting the verses as the Prophet Muhammad did in order to preserve their meanings. The process of traditional Tajweed learning requires a human who gives the learner feedback on their recitation. This supervisor must have a license to allow him to teach Quran. Moreover, a person with a Tajweed license cannot be a supervisor immediately, as he must have a good experience so that he can detect mistakes with high accuracy. For those reasons, the number of Quran Tajweed supervisors is not enough to help all Muslims practice recitation daily and have feedback for every recitation try.

Although the presence of a human teacher in the learning process has its known advantages, it also has disadvantages, so that the automatic learning process can help in overcoming those disadvantages. These systems will be an assistant for teachers at mosques, not a replacement. The goal of such systems is to help the largest number possible of learners to memorize the Holy Quran and correct recitation mistakes daily in a shorter time.

Most of the works in Arabic mispronunciation detection have been done on the Arabic language in general [5]–[10], while only some researchers focused on detecting mispronunciation of Quran recitation. Unfortunately, all of these researchers used private datasets and none of them released their dataset to the public for use by other researchers [11]–[20]. We used a public dataset as we hope that future researchers will compare their results with ours and the automated process will be better and more efficient for real-life application use. Our Experiments were done on a new set of Tajweed rules that have not been studied yet in a research paper as far as we know, with high accuracy in detecting the mispronunciation of it up to 96% achieved by using Long-Short term memory neural network.

# 2. Related Works

Although the limited number of researches on recognizing the recitation of the Holy Quran, this field has gained researchers attention in recent years. Many researches have been conducted to recognize the basic aspects of the recitation, letters, and words [13], [14], [16], [19], or to detect Tajweed rules' mispronunciation [11], [12], [15], [17], [18], [20]. A lot of techniques were used in these papers, some of them are traditional, such as Hidden Markov Model (HMM) [16], [19] and Gaussian Mixture Model (GMM) [14], and some of them are machine learning models such as Support Vector Machine [12], [18], [20] and Multi-Layer Perceptron (MLP) [15], [17], as follows:

Tabbal et al. in [19] developed a delimiter that automatically separates verses from audio files, and an Automatic Speech Recognizer (ASR) for the Arabic language and the recitation of the Holy Quran. The HMM classifier and Mel Frequency Cepstral Coefficient (MFCC) features were used on a private dataset that is one hour of recitations of Surah Al-Ikhlas only. The best accuracy obtained by this system was 85% for females and 90% for males.

Al-Ayyoub et al. in [18] used machine learning to build a system for the automatic recognition of Quran Tajweed rules. This system is able to determine the recitation correctness of the following eight Tajweed rules:

EdgamMeem[4], EkhfaaMeem[5], Tafkheem Lam[6], Tarqeeq Lam[7], Edgam Noon[8] (Noon), Edgam Noon (Meem), Edgam Noon (Waw) and Edgam Noon (Ya'). The authors used a dataset consisting of 3,071 audio files, each containing a recording of exactly one of the eight used rules (in either the correct or the incorrect usage of the rule). Moreover, they compare several feature extraction techniques such as Linear Predictive Code (LPC), Mel-frequency, Cepstral Coefficients (MFCC), Multi-Signal Wavelet Packet Decomposition (WPD), and Convolutional Restricted Boltzmann Machines (CRBM). As for classification, several classifiers were used such as k-Nearest Neighbors (KNN), Support Vector Machines (SVM), Artificial Neural Networks (ANN), and Random Forest (RF), with accuracy of 96% using SVM.

Alagrami et al. in [20] proposed a system that makes use of threshold scoring and support vector machine (SVM) to automatically recognize four different Tajweed rules (Edgham Meem, Ekhfaa Meem, takhfeef Lam, Tarqeeq Lam) with 99% accuracy, where the filter banks were adopted as feature extraction. The dataset used contained about 657 records of Arabic natives and non-natives, each rule has 160 records, and each of them is either the correct pronunciation or the wrong pronunciation of this rule.

A Tajweed classification model was developed by Ahmad et al. in [17]. This solution focused on a set of Tajweed rules called "the Noon Sakinah rules" and in particular the rule of "Idgham" with and without "Ghunnah". Mel-Frequency Cepstral Coefficient (MFCC), and Multilayer Perceptron (MLP) were used for the feature extraction and the classification process, where Gradient Descent with Momentum, Resilient Backpropagation, and the Levenberg-Marquardt optimization algorithm were used to train the neural network. The Levernberg Marquardt algorithm achieved the highest test accuracy (77.7%), followed by Gradient Descent with Momentum (76.7%) and Resilient Backpropagation (73.3%). The dataset used was 300 audio files of recitation of two famous reciters, and each is a recitation of one of those Tajweed rules.

In [16], Rahman et al. proposed an Automated Tajweed Checking System to help children learn the correct recitation of the Holy Quran. Mel-Frequency Cepstral Coefficient (MFCC) was used for feature extraction, and Hidden Markov Model (HMM) for classification and recognition. Using the HMM algorithm, the system can identify and highlight any discrepancy or inconsistency in children's recitation by comparing it with the correct teacher's recitation that is stored in a database, where only one chapter of the Quran was supported, Surat Al-Fatihah.

In [15], Hassan et al. introduced a system to recognize Qalqalah Kubra[9] (القلقلة الكبرى) pronunciation using Multilayer Perceptron (MLP) as classifier and MFCC as features extraction. The dataset used included only 50 samples of the recitation of five different words from the Quran, where for each word there are ten records of correct and incorrect pronunciation, and the achieved results ranged from 95% to 100%.

Putra et al. [14] developed a Speech Recognition System as Quranic Verse Recitation Learning Software. The system used Mel Frequency Cepstral Coefficient (MFCC) for feature extraction and the GMM model as a classifier. In order to test the reliability and accuracy of correction, a data set was collected from ten speakers reading some verses incorrectly and correctly for each of them. The achieved correction accuracy was 90% for hija'iyah letters (Arabic Alphabet Letters) pronunciation, 70% for Tajweed rules ("Idgham", "Ikhfa'" or "Idhar"), and 60% for the combination of pronunciation and Tajweed rules.

---

[4] EdgamMeem : Merge the words through the "Meem" and use "Ghunnah" if the word starts with a "Meem" and is followed by a "Meem Sakinah" [49].

[5] EkhfaaMeem: Apply a "Ghunnah" while concealing the "Meem Sakinah" if a [ب] is followed by a "Meem Sakinah," and then move on to the [ب] [49].

[6] Tafkheem Lam: The word of Allah or Allahum will have heavy "Laam" if there is a "Fatha" (َ) or "Dhamma" (ُ) before it. [49].

[7] Tarqeeq Lam: The "Laam" in Allah or Allahum will be light if there is a "Kasrah" (ِ) before the word. [49].

[8] Edgam Noon: If any of the "Idgham" letters come after the "Noon Sakin" or the "Tanween," the reciter should pronounce the next letter instead of the "Noon" or the "Tanween" [49].

[9] Qalqalah: it is the vibration of sound at the end of the pronunciation of a letter. The letters of Qalqalah: قطب جد [49].

Muhammad et al. [13] proposed the E-Hafiz system to facilitate reciting learning of the holy Quran. Mel Frequency Cepstral Coefficient (MFCC), Vector Quantization (VQ), and Calculation of distance between vectors were used to extract features, reduce the numbers of features vectors and compare the result with the threshold value. A dataset of 10 expert recitations of the first 5 surahs of the Holy Quran was used and recognition accuracy of 92%, 90%, and 86% was achieved for men, children, and women, respectively.

Nahar et al. in [12] took a different path as they proposed a recognition model to recognize the “Qira’ah” from the Holy Quran recitation precisely 96%, since according to the narration "hadith" No. 5041, taken from [21], the Holy Quran has seven main reading modes, known as "Qira'at," which are acknowledged as the most popular methods of reciting from the Holy Quran, and three complementary readings of the seven. This model used the Mel-Frequency Cepstrum Coefficients (MFCC) features and Support Vector Machine (SVM), where the authors have built a dataset has 10 categories, each one representing a type of Holy Quran recitation or “Qira’ah”, with a total of 258 wave files.

**Table 1** Comparison between researchers

| Work | Tajweed Rules | Dataset | Feature Extraction | Model | Results |
|---|---|---|---|---|---|
| [19] | Sourate Al-Ikhlass with the most important Tajweed rules | About 1 hour of audio recitations of Sourate Al-Ikhlass for different reciters | Mel-Frequency Cepstral Coefficient (MFCC) | Hidden Markov Model (HMM) based on Sphinx. | 85% - 90% |
| [15] | One Tajweed rule (Qalqalah) | 100 samples for Qalqalah pronunciation | Mel-Frequency Cepstral Coefficient (MFCC) | Multi-Layer Perceptron (MLP) | 95% - 100% |
| [11] | Check the Makhraj (place of utterance) of Arabic phonemes and allophones. | Ten combinations of alphabets from 10 users, resulting in 20,000 samples. | Mel-Frequency Cepstral Coefficient (MFCC) | Mean Square Error (MSE) | Up to 100% accurate for a one-to-one mode |
| [18] | Eight Tajweed rules (EdgamMeem,EkhfaaMeem, Tafkheem Lam, Tarqeeq Lam, Edgam Noon (Noon), Edgam Noon (Meem), Edgam Noon (Waw) and Edgam Noon (Ya’)) | 3071 audio files, each containing a recording of exactly one of the eight rules | Convolutional Restricted Boltzmann Machines (CRBM), MFCC, WPD, HMM-SPL | Support Vector Machine (SVM) | 96.4% |
| [20] | Four Tajweed rules (Edgham Meem, Ekhfaa Meem, takhfeef Lam, Tarqeeq Lam) | 657 recordings of 4 different rules total, | Mel-Frequency Cepstral | Support Vector Machine (SVM) | 99% |
| | | with about 160 records for each rule | Coefficient (MFCC) | | |
| [17] | Rule of “Idgham” with and without “Ghunnah” | 300 audio files | Mel-Frequency Cepstral Coefficient (MFCC) | Multi-Layer Perceptron (MLP) | 77.7% |
| [16] | Surat Al-Fatihah. | One Correct recitation of Surat Al-Fatihah. | Mel-Frequency Cepstral Coefficient (MFCC) | Hidden Markov Model (HMM) | N/A |
| [13] | Word level for verses in the dataset | Consists of 10 expert recitations of the first 5 surahs of the Holy Quran. | Mel-Frequency Cepstral Coefficient (MFCC) | Threshold based on Euclidean distance | 86% - 92% |
| [14] | Letter level and some Tajweed rules | A voice recording of a Quran recitation expert | Mel-Frequency Cepstral Coefficient (MFCC) | Gaussian Mixture Model (GMM) | 60% - 90% |
| [12] | Classifying the recitation into one of the ten “Qira'at” | 258 wave files | Mel-Frequency Cepstral Coefficient (MFCC) | Support Vector Machine (SVM) | 99% |

# 3. Methodology

In this paper, we proposed a deep learning model to detect mispronunciation of a new set of Tajweed rules using Long-Short Term Memory (LSTM) neural networks. The digital speech signal is treated as a series of audio frames. Each frame consists of a set of samples [22], hence the mistake in pronunciation can appear on several consecutive frames. LSTM is a suitable choice for detecting mispronunciation as it has internal memory and can take advantage of this sequence to make the right decision [23] better than the other researches where decisions were made without properly considering the sequence of frames [12], [15], [17], [18], [20].

Our proposed solution schema is described in Figure 1 where the speech signal is pre-emphasized and features are extracted from it then these features are entered into the deep learning model, and finally the pronunciation of the Tajweed rule is judged as true or false. The following subsections contain a detailed explanation of each step.

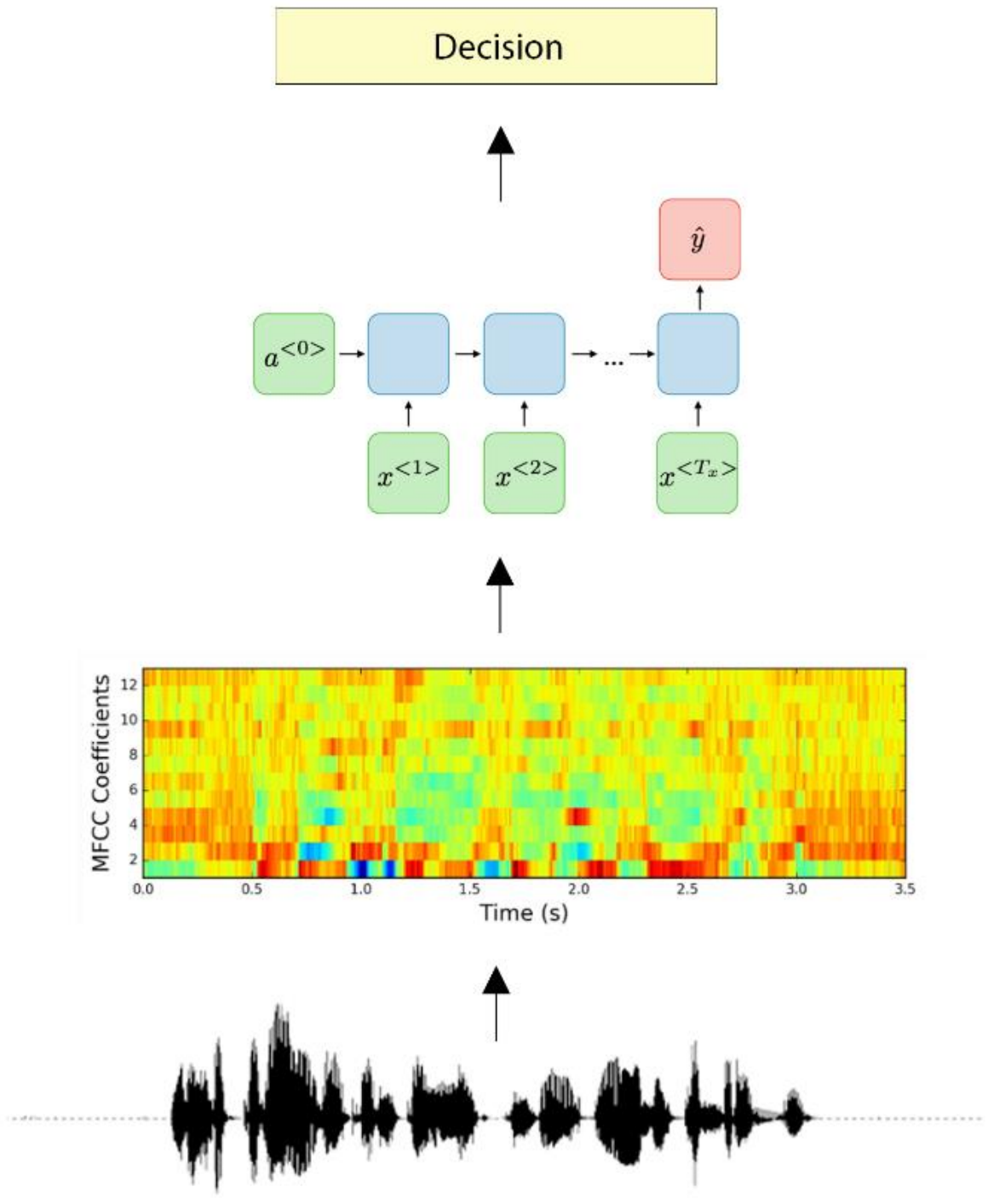

**Fig. 1** Our Proposed Solution Schema

## 3.1 Dataset

Although many researches were done to recognize Tajweed rules on Holy Quran, none of them made their dataset public to be used by other researchers. In this work, we have used a public dataset, QDAT [24] which contains more than 1500 voices of the correct and incorrect recitation of three Tajweed rules: Separate stretching[10], Tight Noon[11], and Hide[12]. To the best of our knowledge, nobody else has previously worked on yet. The audio clips were recorded by people of both genders and of different age groups and included recording a part of verse 109 of Surah AlMa'idah. It is recorded using "WhatsApp" online in WAV format with an 11kHz sample rate, mono channel, and 16-bit resolution.

## 3.2 Pre-emphasis

Pre-emphasis is a signal processing technique that addresses the attenuation of high frequencies generated during the human speech production process by increasing the amplitude of the high-frequency bands and decreasing the amplitude of the low-frequency bands. This technology helps acoustic models better detect phones, where it lessens the sensitivity of these models to noise [25]. The first-order Finite Impulse Response (FIR) filter from Equation 1 is the pre-emphasis filter that is most frequently used.

$$H(z) = 1 - \alpha z^{-1} \tag{1}$$

[10] The separate stretching: is that the letter of the dowel in the last word, and the Hamza is the first word that follows it. The duration is four or five movements.

[11] Tight Noon: Ghunnah must be shown in the aggravated Noon by two movements.

[12] Hide: It is pronouncing the consonant noun or Tanween in a state between showing and fading without emphasis, with the song remaining by two movements.

Where $\alpha$, is the factor of pre-emphasis and is equal to 0.97 in this work.

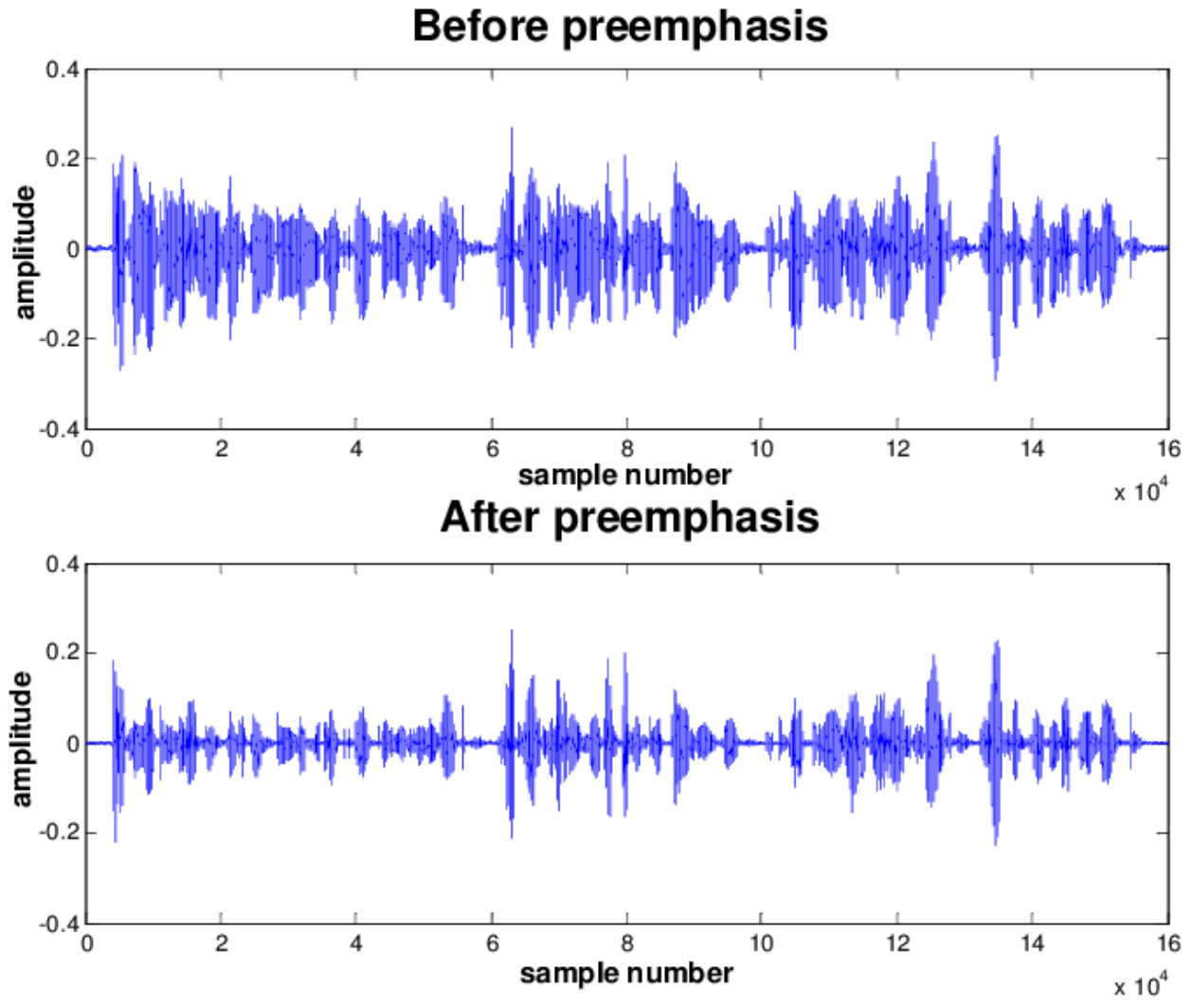

**Fig. 2** Signal before and after emphasis [26]

## 3.3 Features Extraction

Calculating Mel-Frequency Cepstral Coefficients (MFCC) is the most widespread and major method for extracting spectral features in speech recognition applications [27], whereby, it is based on the Fourier Transform and the frequency domain of the Mel scale to mimic the human ear scale [28]. Tewari [29] identified the procedure for extracting the MFCC features, as illustrated in Figure 3.

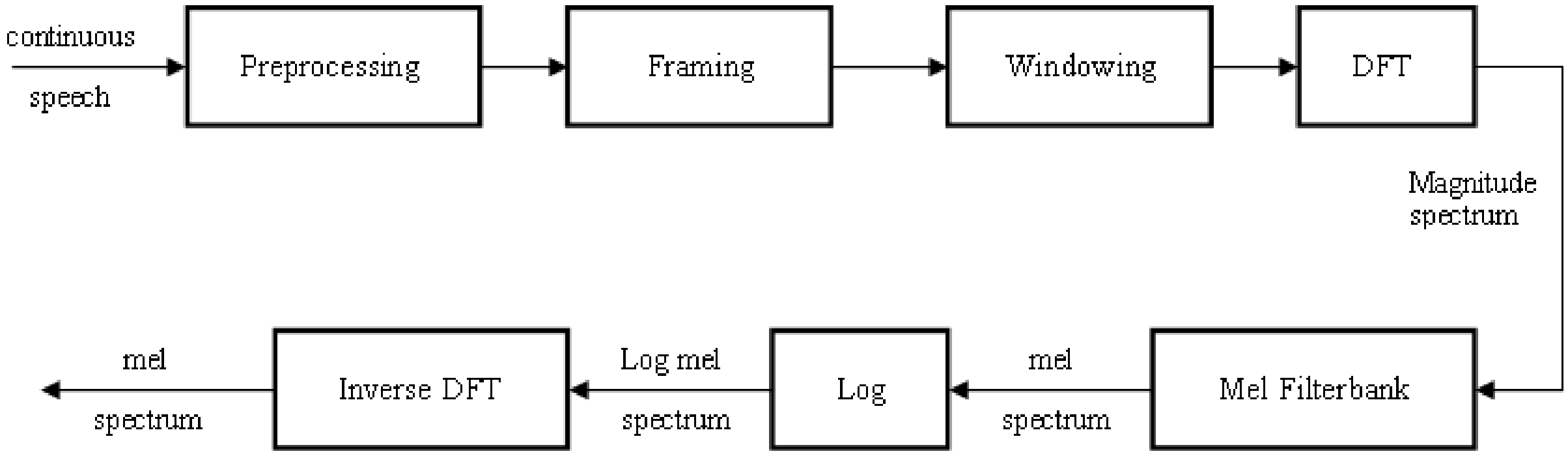

**Fig. 3** Block diagram of the computation steps of MFCC [29]

### 3.3.1 Framing

Framing is the division of voice into multiple frames with a duration of between 20 and 40 msec, where each frame consisted of several samples and can be represented by a single vector. This procedure is required to obtain stable acoustic properties for speech analysis, by turning the non-stationary signal into a quasi-stationary signal, which can then undergo the Fourier transformation to move from the time domain to the frequency domain [30].

### 3.3.2 Windowing

The windowing process is used to reduce and get rid of the unexpected fall at the beginning and end of each frame of the signal, which produces noise in the high-frequency domain while taking the Discrete Fourier transformation on the signal. One of the most famous functions is Hamming window [25] which was used in this work and given by Equation 2 as follows: N is the window length, and n is the $n^{th}$ sample in the frame.

Where 0 ≤ n ≤ N-1:

$$W_n = \begin{cases} 0.52 - 0.46\cos(\frac{2\pi n}{N-1}), & 0 \le n \le N-1 \\ 0 \quad , & others \end{cases} \tag{2}$$

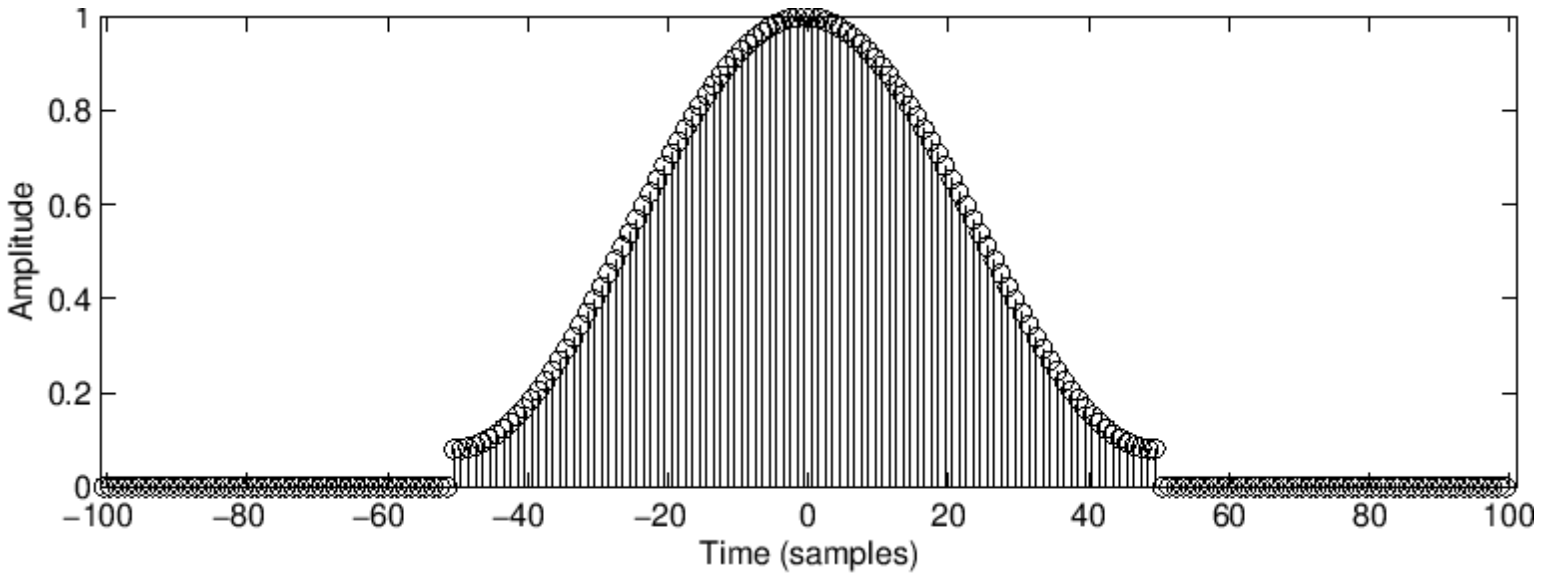

**Fig. 4** Hamming window [31]

### 3.3.3 Discrete Fourier Transform (DFT)

Discrete Fourier transform is performed for each frame to calculate the frequency spectrum by converting signals from the time domain to the frequency domain. The Fast Fourier Transform (FFT) algorithm is usually used to calculate DFT instead of using DFT directly [32] because its complexity is less than that of the DFT. As shown in Equation 3, the input of DFT is a windowed signal, $Y1[k]$, and the output is a complex number, $Y2[n]$, which represents the magnitude and the phase of that frequency component.

$$S[n] = \sum_{k=0}^{N-1} \mathrm{Y}[k] e^{-2\pi jkn/N} \tag{3}$$

Where $N$ is the number of samples, $k = 0, 1, 2, 3, \ldots, N-1$ and $S[n]$ is the Fourier Transform of $Y[k]$.

### 3.3.4 Mel Filter-bank

Compared to high frequencies, humans are more sensitive to the low-frequency component of speech and can get more useful information through it. The human auditory system perceives the physical frequency of the tone in a form that is more logarithmic than linear, and this is what the Mel scale does. It shows how the frequency in Hz and in Mel scale are related to each other, where it is approximately a linear frequency spacing below 1 kHz and a logarithmic spacing above 1 kHz [33]. This process is done by using a number of triangular filter banks, the most commonly used shape, which their width increases as the frequency gets higher. The $Hz$ to $Mel$ mapping is expressed in Equation 4 [34]:

$$Frequency\ (Mel\ Scale) = 2592 * log10\ (1 + f/700) \tag{4}$$

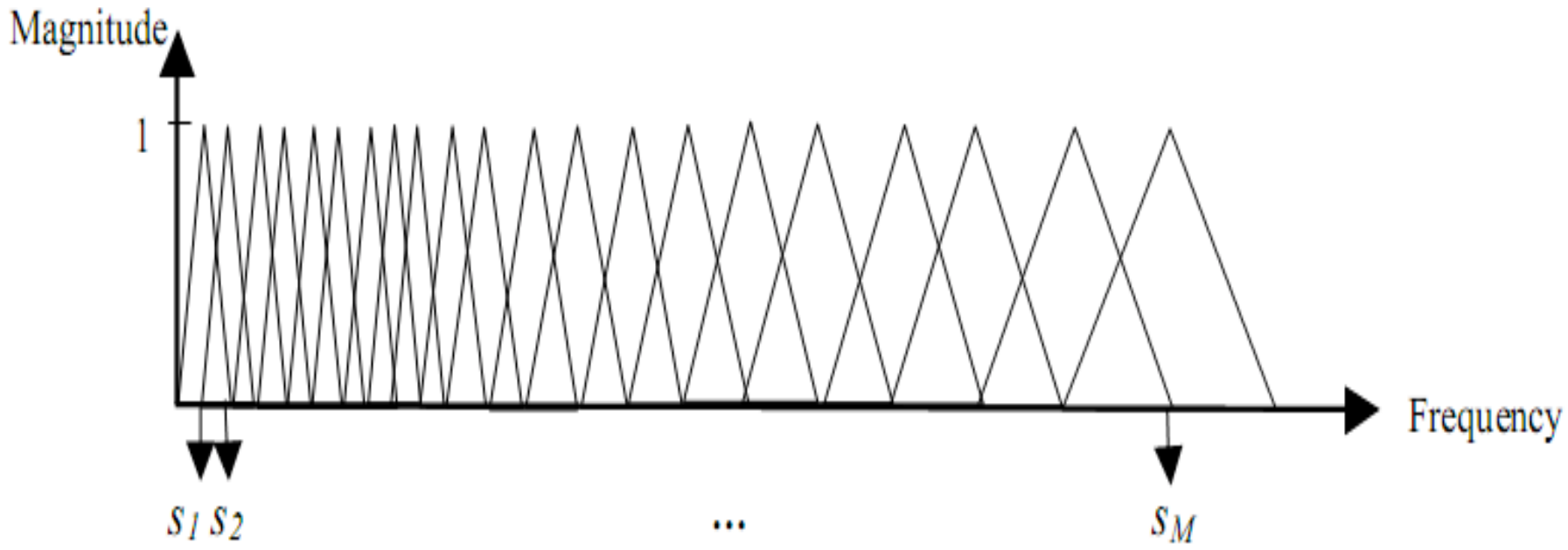

**Fig. 5** Mel filter bank construction [35]

### 3.3.5 Logarithm

Just as the human hearing system is more sensitive to frequency differences at lower frequencies, it is also more sensitive to energy differences at lower-energy signals than high-energy signals, where the human response to the signal level is logarithmic [36]. For that, the log function is applied to the Mel filter bank's output, to mimic humans' hearing.

### 3.3.6 Inverse of Discrete Fourier Transform (IDFT)

The inverse DFT is applied to the output energies from the previous steps since they are correlated due to the overlapping of the filter banks, and it is the last step to obtain the MFCC features. This step produces feature vectors obtained by Equation 5 [25], where $Y[n]$ are the cepstral coefficients, $x[k]$ represents the logged value of each speech segment produced by the Mel filter in the previous step, and N is the number of MFCC coefficients.

$$Y[n] = \sum_{k=0}^{N-1} x[k] \cos\left[\frac{n \langle k - \frac{1}{2} \rangle \pi}{N}\right], n = 1,2,3 \ldots, N \qquad (5)$$

## 3.4 Long-Short Term Memory (LSTM)

A recurrent neural network (RNN) is a type of artificial neural network, introduced by Rumelhart et al. [37], which can handle sequential data or time series data. RNNs are often used for ordinal or temporal problems, and what sets them apart is their "memory" which allows them to use data from earlier inputs to affect the input and output at hand. Unlike traditional deep neural networks, which consider inputs and outputs to be independent of one another. RNN suffers from vanishing gradient [38] and exploding gradient [39] problems and so Long Short-Term Memory networks (LSTM) were introduced by Hochreiter & Schmidhuber [23] that overcomes these problems and can learn long-term dependencies. LSTM have been used in complex problem domains such as language translation, natural language processing, speech recognition, image captioning and more [40]–[43].

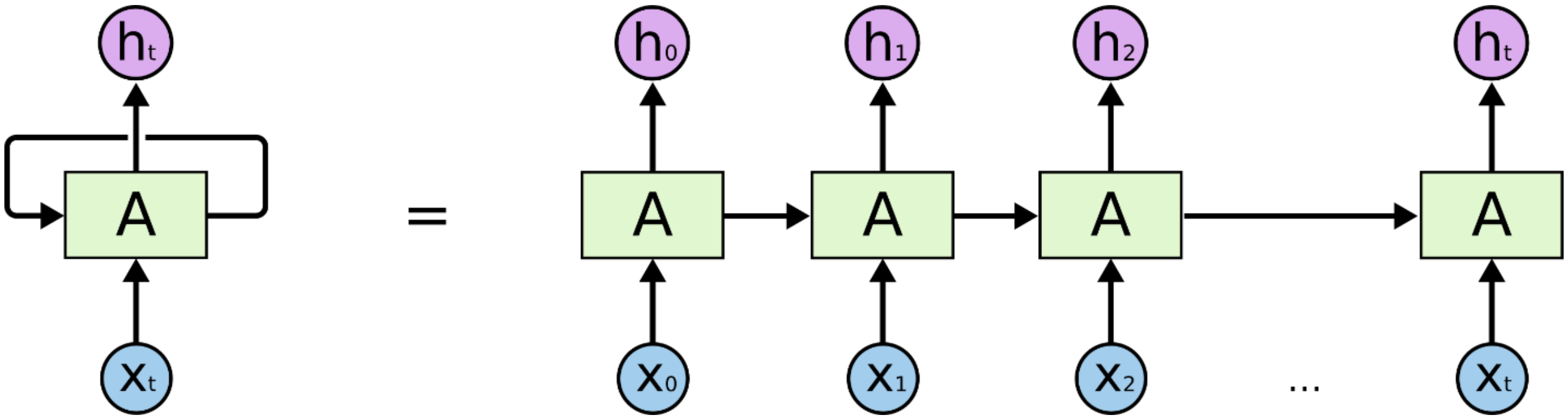

**Fig. 6** Recurrent Neural Network Structure [44]

The LSTM's central idea is the cell state, which works as the memory of the network by carrying relevant information as the sequence is processed. Information is added or deleted from the cell state by neural networks called gates, which consist of a pointwise multiplication operation and a layer of the sigmoid neural network [23].

Figure 7 shows LSTM cell architecture, where the cell state $C_t$ at time $t$ and the three main gates input gate $i_t$, forget gate $f_t$ and output gate $o_t$ make up each LSTM cell.

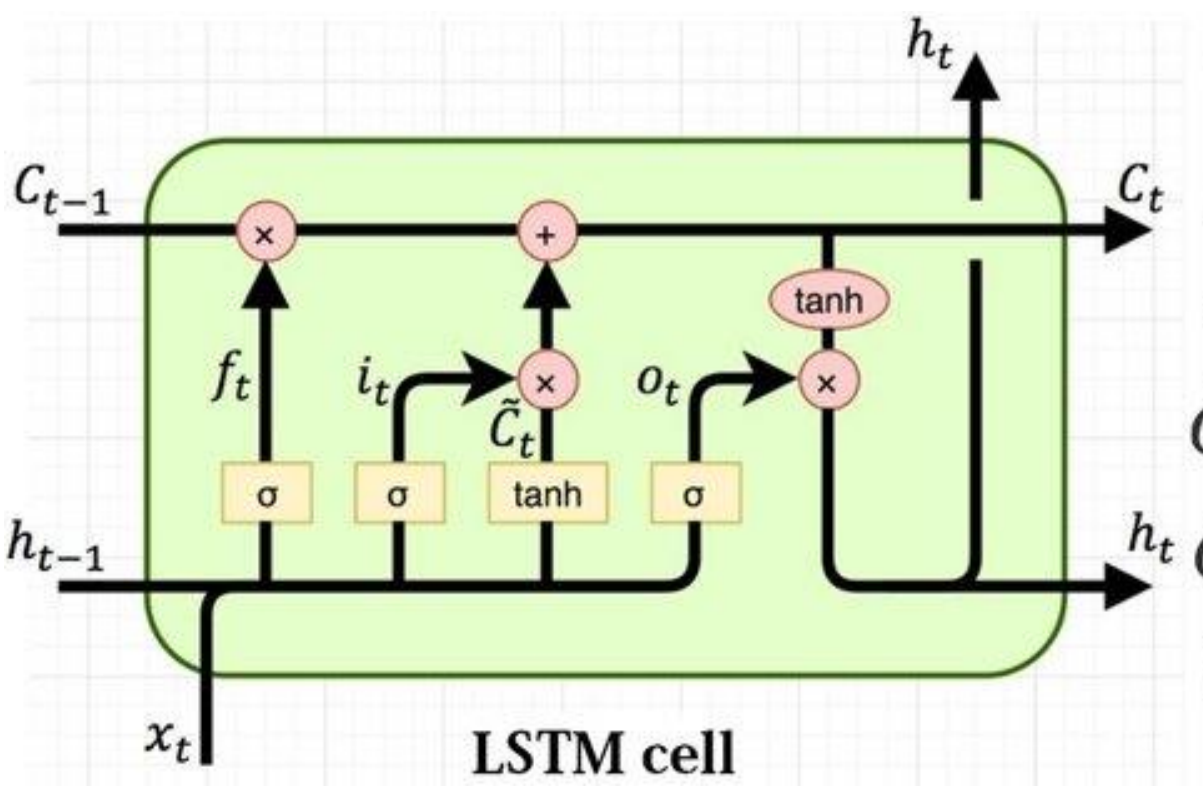

$$i_t = \sigma\left(x_t U^i + h_{t-1} W^i\right)$$
$$f_t = \sigma\left(x_t U^f + h_{t-1} W^f\right)$$
$$o_t = \sigma\left(x_t U^o + h_{t-1} W^o\right)$$
$$\tilde{C}_t = \tanh\left(x_t U^g + h_{t-1} W^g\right)$$
$$C_t = \sigma\left(f_t * C_{t-1} + i_t * \tilde{C}_t\right)$$
$$h_t = \tanh(C_t) * o_t$$

**Fig. 7** Long-Short Term Memory (LSTM) architecture and equations [45]

### 3.5 Experiments setup

To extract MFCC features, we applied the Hamming window function on frames of 32ms and used 40 Mel filter banks. These values were selected after many experiments while tuning the parameters. The dataset was divided into training and testing sets as 90% and 10%, respectively. The training set was used to train the deep learning model which consists of three LSTM layers, each has 256-unit, in addition to three fully connected layers with 16, 16, and 8 units, respectively, each using the RELU function as the activation function, two Dropout layers with a dropout rate of 0.2, and finally one fully connected layer for each rule as a classification layer which uses sigmoid function as an activation function. The binary entropy loss function and the RMSprop optimizer with a learning rate of 0.0005 were also used in this model.

## 4. Results & Discussion

We compared the results of our deep learning model with the results of traditional machine learning methods proposed by previous works on the QDAT dataset, as they used private datasets.

**Table 2** The accuracy achieved through deep learning and machine learning models

| | Separate Stretching | Hide | Tight Noon |
|---|---|---|---|
| Random Forest | 91% | 83% | 91% |
| KNN | 91% | 81% | 94% |
| SVM | 81% | 71% | 85% |
| Logistic Regression | 88% | 79% | 91% |
| **LSTM (Our Model)** | **96%** | **95%** | **96%** |

We can justify these results from the fact that using deep learning in complex problems such as speech recognition is better than using traditional machine learning algorithms [46] as well as that deep learning gives better results when the size of training data increases [47][48]. Also, the use of LSTM networks, which clearly outperformed the other works, had the largest impact on accuracy, as it takes into consideration the sequence of acoustic frames better than the traditional machine learning algorithms used by other researchers.

**Table 3** Details of the results

| | Separate Stretching | Hide | Tight Noon |
|---|---|---|---|
| **Accuracy** | 96% | 95% | 96% |
| **Recall** | 95% | 95% | 98% |
| **F1-score** | 95% | 95% | 97% |

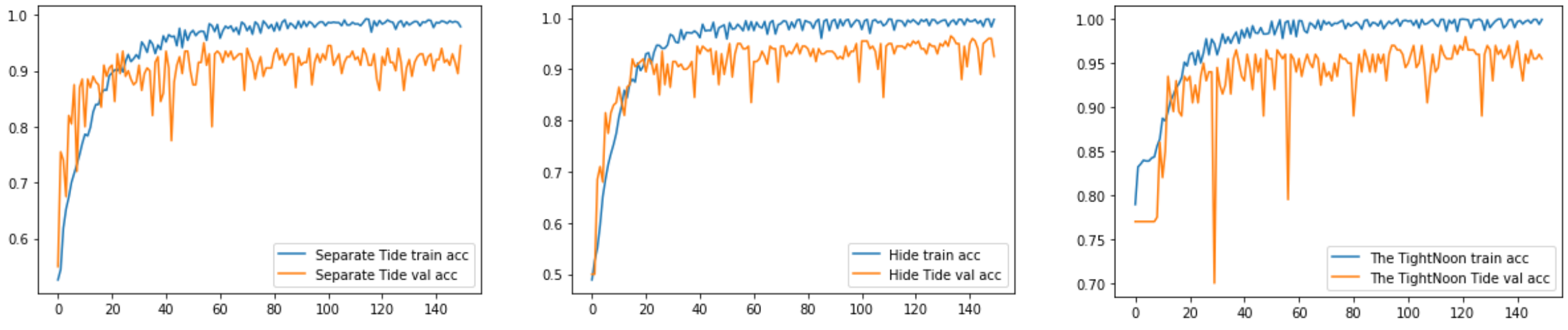

**Fig. 8** Accuracy curves

# 5. Conclusion

Correct pronunciation of Tajweed rules is necessary to read the Holy Quran. The automatic detection of pronunciation mistakes in these rules is an area that has not received much attention by researchers despite its necessity due to the insufficient number of human supervisors for traditional Tajweed learning methods. For those reasons, many papers have been done on automatic Tajweed learning. However, all previous works have three common problems, first, they focused on machine learning only. Second, they used private datasets with no benchmark to compare with. Third, they did not take into consideration the sequence of input data optimally, although the speech signal is a series of frames. To overcome these problems, we proposed a solution that consists of Mel-Frequency Cepstral Coefficient (MFCC) features with Long Short-Term Memory (LSTM) neural networks which use the time series, to detect mispronunciation in Tajweed rules. In addition, our experiments were performed on a public dataset, the QDAT dataset. We compared our results with traditional machine learning algorithms from previous works where our solution outperforms them. Our proposed solution showed excellent results that reached 96% for all rules.

# 6. Declarations

## Authors' contributions

AAH performed the literature review, conducted the experiments, and wrote the manuscript.
KAJ took on a supervisory role and made a contribution to the conception and analysis of the work.
All authors read and approved the final manuscript.

## Funding

The authors declare that they have no funding.

## Data Availability

The data set used in this work is available at:

https://www.kaggle.com/datasets/annealdahi/quran-recitation

## Ethics approval and consent to participate

The authors Ethics approval and consent to participate.

## Consent for publication

The authors consent for publication.

## Conflicts of Interest

The authors declare that they have no competing interests.

## Acknowledgments

I would like to thank my parents for their support.